# THE GRACEFUL EXIT PROBLEM
# IN
# STRING COSMOLOGY

Ram Brustein

Theory Division, CERN

CH-1211 Geneva 23, Switzerland

and

G. Veneziano

Theory Division, CERN

CH-1211 Geneva 23, Switzerland

and

Observatoire de Paris, DEMIRM, 75014, Paris, France

## ABSTRACT

Isotropic string cosmology solutions can be classified into two duality related branches: one can be connected smoothly to an expanding Friedman-Robertson-Walker Universe, the other describes accelerated inflation or contraction. We show that, if the dilaton potential has certain generic properties, the two branches can evolve smoothly into each other. We find, however, that just the effects of a potential do not allow for a graceful exit from accelerated inflation in the weak curvature regime. We explain how conformal field theory techniques could be used to give a decisive answer on whether graceful exit from accelerated inflation is possible at all.

There has been growing interest, recently, in the question of whether or not inflationary cosmology can be accommodated naturally in superstring theory and arguments in both directions have been presented.

On one hand, dilaton interactions appear to inhibit inflationary evolution [1],[2]. This observation is based on the standard assumption that inflationary evolution is driven by a non-vanishing potential energy density. Since the dilaton couples necessarily to any kind of energy density, a dilaton potential is induced. The induced potential turns out to be so steep that the dilaton kinetic energy dominates over the potential energy leaving nothing to drive super-luminal expansion. Some attempts have been made to overcome this fact by assuming that inflation occurs after the dilaton has settled down to its minimum or invoking anthropic-like arguments about the initial conditions of the dilaton [3]. Neither of these seem to be completely satisfactory.

On the other hand, the duality-like symmetries that characterize the equations of string cosmology [4], [5] suggest a totally different mechanism [4], [6] for inflationary evolution. Rather than relying on potential energy as source for inflation it involves a novel and interesting interplay between the dilaton and the metric. This mechanism is based on the fact that cosmological solutions to string dilaton-gravity come in duality-related pairs (branches) connected to a sign ambiguity in solving a quadratic equation. We will thus call them the plus, (+), and the minus, (−), branch. The (−) branch can be connected smoothly to a standard Friedman−Robertson−Walker (FRW) decelerated expansion of the universe with constant dilaton, while the (+) branch, even in the absence of a dilaton potential, has solutions of the accelerated inflation type (so called superinflation [7]), in which the Hubble parameter increases with time. Duality related cosmologies are actually part of a larger, continuous family of $O(d,d)$ related solutions [8] which involve the antisymmetric tensor as well. For the purpose of this paper, this larger family does not appear to play an important role.

The mere existence of duality related solutions does not provide per se an answer to the questions raised in [1] and [2]. The real question, that we set out to answer, is whether or not the dynamics itself likes to pair duality-related solutions in a single, continuous history of the universe taking place in cosmic time. In other words, we would like to understand whether string cosmology allows the inflationary phase to end and to connect smoothly to a FRW expanding universe[1]. So far such a picture was seen to emerge either by invoking an ad-hoc symmetry principle (self-duality) or by using quite contrived forms of dilaton potentials [6]. Neither of these appears to be completely satisfactory.

Here we use the low energy effective field theory approach to set up the problem. We assume therefore, throughout our analysis, that curvatures and derivatives are small in string units. We also assume that quantum corrections are not too large, i.e., we work in a region where string theory is weakly coupled. This is consistent with the standard view of string theory whereby the gauge coupling constant at the string scale is related, in our conventions, to the string-loop expansion parameter by $e^{\phi} = g_{string}^2 \sim g_{gauge}^2$. The present value of the dilaton is therefore reasonably perturbative and we shall assume that it had never been much larger than that in the past. As we shall see below, this fits very well with the whole scenario.

The transition between inflationary evolution and a FRW expanding universe is usually referred to as "graceful exit" from inflation and it is a well known problem to be faced by any model of inflation. The possibility of graceful exit from accelerated inflation is closely related, in our setup, to the question of whether the two branches can be smoothly connected to one another. Incidentally, the notion that small curvature solutions with different characteristics (e.g. topology) can be smoothly connected to one another as some parameter is varied continuously is not new in string theory [10]. What is new in our context is that the parameter should represent physical (cosmic) time. If such "branch changes" can occur dynamically, the following appealing scenario can be easily imagined. The dilaton starts somewhere in the weak coupling region and, initially, the Hubble parameter, $H$, is also very small. Both then evolve as in the (+) branch solution. The universe inflates and the dilaton evolves toward stronger coupling. This evolution is essentially unaffected by the dilaton potential or by matter or radiation sources. At some later time a "branch change" occurs. The potential and sources become important, the universe evolves as a regular FRW

---

[1]In pure Brans-Dicke theory a related analysis was carried out by Weinberg [9]

Universe, reheats and the dilaton settles down at a minimum of its potential.

We will find that smooth branch changes do occur if the dilaton potential has certain generic properties. We also find, however, that, it is not possible to find a solution which describes a complete, graceful exit from accelerated inflation. The analysis points to the possibility that strong curvature effects may play an essential role in allowing graceful exit from accelerated inflation. Analysis of exact (to all orders in $\alpha'$) stringy solutions should give a definite answer concerning the viability of the whole idea and encouraging results in this direction have been recently obtained [11]

There are different "frames" in which one can describe Brans-Dicke (BD) type theories. These are related by local field redefinitions which, supposedly, do not affect physical observables (see e.g. [12]). Here we perform the analysis in the BD (string) frame because the physics interpretation is simpler in such frame [13],[2] and the arguments are more transparent.

The relevant terms in the superstring effective action below the Planck scale are:

$$S_{eff} = \frac{1}{16\pi\alpha'} \int d^4x \sqrt{-g} \left\{ e^{-\phi} \left[ R + \partial_\mu \phi \partial^\mu \phi \right] - V(\phi) + \cdots \right\} \quad (1)$$

where we have assumed to be in the critical number of dimensions (central charge $c = 10$ with the aid of some passive coordinates or internal degrees of freedom) while all the interesting dynamics takes place in ordinary 4-dimensional space-time.

Our conventions are those of Weinberg [9]. Note that the dilaton ($\phi$) kinetic term has the "wrong" sign, corresponding to a BD $\omega$ parameter equal to $-1$, as required by duality [4]. The origin of the dilaton potential $V(\phi)$ is expected to be non-perturbative, related to supersymmetry breaking, and derived from the superpotential of the theory in the standard way. There will be also perturbative and non-perturbative corrections to the coefficients of the other terms in the action, but they are expected to be small because of the assumption that the theory remains weakly coupled throughout the cosmological evolution. As will become clear, the details of the shape of the potential are not particularly important for our analysis. Finally, the dots stand for terms involving other fields, more derivatives etc.

The equations of motion resulting from the action (1) are

$$R_{\mu\nu} - \frac{1}{2}g_{\mu\nu}R - \frac{1}{2}g_{\mu\nu}\partial_\mu\phi\partial^\mu\phi + g_{\mu\nu}D_\mu D^\mu\phi - D_\mu D_\nu\phi - \frac{1}{2}g_{\mu\nu}U(\phi) = 0$$
$$R + D_\mu\phi D^\mu\phi - 2D_\mu D_\mu\phi - U(\phi) = 0 \quad (2)$$

where $U(\phi) \equiv e^\phi V(\phi)$. We look for solutions of the above equation in which the metric is of the isotropic, FRW type with vanishing spatial curvature (we will discuss later more general cases) and the dilaton depends only on time.

$$ds^2 = -dt^2 + a^2(t)dx_i dx^i$$
$$\phi = \phi(t) \quad (3)$$

The Hubble parameter, $H$, is related to the scale factor, $a$ in the usual way, $H \equiv \frac{\dot{a}}{a}$.

After some algebra one obtains two independent first order equations for the dilaton and $H$ [4]. The original dilaton equation is a consequence of these two equations, which read

$$\dot{H} = \pm H\sqrt{3H^2 + U} - \frac{1}{2}U'$$
$$\dot{\phi} = 3H \pm \sqrt{3H^2 + U} \quad (4)$$

The ($\pm$) signifies that either a ($+$) or ($-$) is chosen for both equations simultaneously. These equations are the starting point of our analysis.

The solutions to the equations (4) belong to two branches, according to which sign is chosen. Another twofold ambiguity is related to the sign of $H$. For simplicity, we start by discussing the solutions in the very simple case in which there is no potential. The four solution one obtains in this case are related to one another by scale-factor-duality, time reversal, or both. The ($-$) branch is a regular, negative feedback, branch. The solution $\{H^{(-)}, \phi^{(-)}\}$ as functions of time is simply given by

$$H^{(-)} = \pm \frac{1}{\sqrt{3}}\frac{1}{t - t_0}$$
$$\phi^{(-)} = \phi_0 + (\pm\sqrt{3} - 1)\ln(t - t_0) \ , \ t > t_0 \quad (5)$$

This solution describes decelerated expansion ($H > 0, \dot{H} < 0$) or decelerated contraction ($H > 0, \dot{H} < 0$) depending on the initial sign of $H$. Correspondingly, the evolution is towards strong or weak coupling, respectively. As we show explicitly later (see Eqs.(12) below) this branch can be joined smoothly to an ordinary FRW radiation-dominated expanding universe. Provided the dilaton potential has a minimum with vanishing cosmological constant and suitable sources are added, the dilaton will eventually settle down at it vacuum expectation value and standard cosmology will follow.

The $(+)$ branch has instead some unusual and quite remarkable properties [4],[6]. In the absence of any potential the solution for $\{H, \phi\}$ is now given by

$$
\begin{aligned}
H^{(+)} &= \pm \frac{1}{\sqrt{3}} \frac{1}{t - t_0} \\
\phi^{(+)} &= \phi_0 + (\pm\sqrt{3} - 1)\ln(t_0 - t) \ , \ \ t < t_0
\end{aligned}
\tag{6}
$$

This solution describes either accelerated expansion and evolution from a cold, flat and weakly coupled universe towards a hot, curved and strongly coupled one [6] or accelerated contraction and evolution towards weak coupling. In general, the effects of a potential on this branch are quite mild. The dilaton zooms through potential minima. It is impossible, in this branch, for the dilaton to sit at a minimum of its potential. Inflation, in this solution, is driven, not hampered, by the dilaton's kinetic term, thanks to the negative value $(-1)$ of the BD $\omega$ parameter.

The following table should help clarifying the difference between the two branches

|       | $(+)$ branch            | $(-)$ branch            |
|-------|-------------------------|-------------------------|
| $H < 0$ | $\dot{H} < 0, \ \dot{\phi} < 0$ | $\dot{H} > 0, \ \dot{\phi} < 0$ |
| $H > 0$ | $\dot{H} > 0, \ \dot{\phi} > 0$ | $\dot{H} < 0, \ \dot{\phi} > 0$ |

Table 1. Classification of two branches of solutions as functions of time.

The above analysis becomes quite complicated when one considers non-vanishing potential and matter contributions. Since Eqs.(4) are first order equations we found that it is best to represent their solutions as trajectories in the phase space $\{H, \phi\}$. In Figure 1. some $(+)$ and $(-)$ trajectories are shown. The units on the vertical $H$ axis are arbitrary.

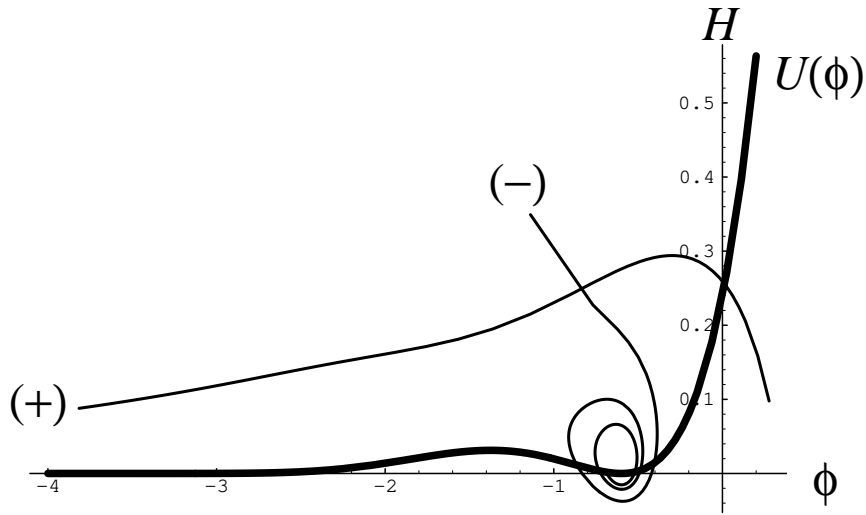

Figure 1. The potential $U(\phi)$ and trajectories of $(+)$ and $(-)$ branches.
The $(-)$ branch settles down at a minimum of the potential.

From Eqs.(4) we see that a branch change can occur only if the potential $V(\phi)$ becomes negative for some range of $\phi$. This is because the only points allowing a continuous transition from one branch to the other are those where

$$3H^2 + U(\phi) = 0 \ , \tag{7}$$

or else there will be a discontinuity in $\dot\phi$. Similarly, continuity in $H$ will imply that, of the four solutions of Table 1, only those with the same sign of $H$ can turn into one another. It is clear that these correspond to pairs of solutions related by the product of scale-factor-duality and time reversal, since the combination of the two operations does not reverse the sign of $H$.

Since $H^2 > 0$ we have to have $U(\phi) < 0$ to allow for solutions to Eq.(7) to exist. The locus of points satisfying $3H^2 + U(\phi)$ is a curve in phase space. For potentials for which $V(\phi) < 0$ in some bounded interval the shape of the curve resembles an egg and we will therefore refer to it from now on as the "Egg". We assume that the "Egg" lies in the mild-coupling region. In Figure 2 a typical dilaton potential and the corresponding "Egg" are shown. In supergravity theories it is a generic property that the global minimum of the potential is negative. Moreover, if one considers a generic potential induced by field theoretic non-perturbative effects in superstring theory, it does indeed have the property that it has an Egg and the Egg is completely in the weak (mild) coupling region (see e.g. [1]).

It turns out that a negative potential is not a sufficient condition for inducing a branch change. We turn now to explain the additional condition that ensures that a branch change does indeed occur. Consider a trajectory in phase space near the Egg. We can compute its direction in phase space, $\frac{dH}{d\phi}$ as compared to the direction of the Egg, $\frac{dH}{d\phi}\big|_{Egg}$, for the two branches

$$\frac{dH}{d\phi} - \frac{dH}{d\phi}\bigg|_{Egg} = \pm \frac{1}{3}\epsilon \left[1 - \frac{1}{2}\frac{U'}{U}\right] + \mathcal{O}(\epsilon^2) \qquad (8)$$

where the prime denotes derivative with respect to $\phi$ and $\epsilon = \sqrt{3H^2 + U(\phi)}$ is a measure of the distance of the trajectory from the Egg. Note that $\epsilon > 0$ always. Consider the $(+)$ branch and $H > 0$ so the trajectory is above the Egg as in Figure 2. If $U'$ is negative and large such that the quantity $\left[1 - \frac{1}{2}\frac{U'}{U}\right]$ is negative (recall $U < 0$) the trajectory is attracted to the Egg, touches it and then has to turn into the $(-)$ branch by continuity. If the quantity $\left[1 - \frac{1}{2}\frac{U'}{U}\right]$ is positive the trajectory is repelled from the Egg and a $(+) \to (-)$ branch change will not occur. Consider now the $(-)$ branch and $H > 0$ so the trajectory is above the Egg. If the quantity $\left[1 - \frac{1}{2}\frac{U'}{U}\right]$ is positive the trajectory is attracted to the Egg, then touches the Egg and then has to turn into the $(+)$ branch by continuity. If the quantity $\left[1 - \frac{1}{2}\frac{U'}{U}\right]$ is negative the trajectory is repelled from the Egg and a $(-) \to (+)$ branch change will not occur. When the trajectory is below the Egg, a similar analysis can be carried through with the result that branch changes indeed occur, depending on the value of $\left[1 - \frac{1}{2}\frac{U'}{U}\right]$.

In Figure 2. a $(+) \to (-)$ branch change induced by a negative potential is shown. The units on the vertical $H$ axis are arbitrary.

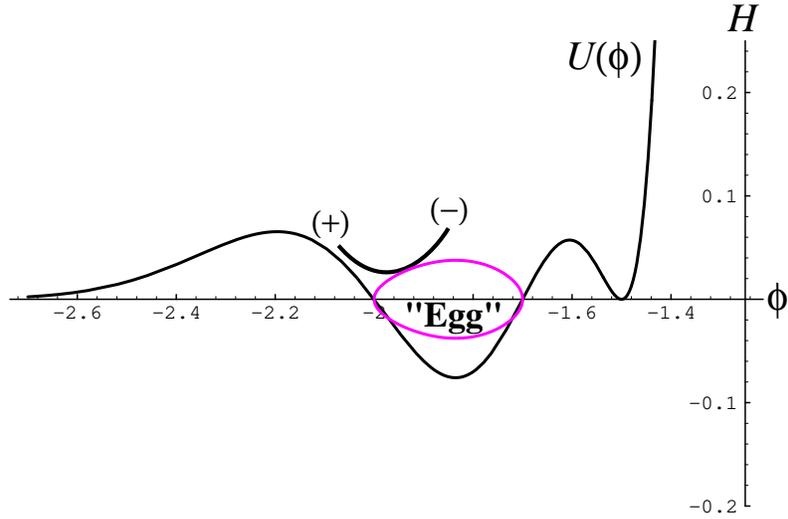

Figure 2. The dilaton potential and corresponding "Egg".
A branch change induced by negative potential.

To our surprise we discovered that, although branch changes can be induced by a negative potential, it does not provide a useful way of achieving a graceful exit from accelerated inflation because branch changes occur in pairs. Let us start explaining this point using the following example. Suppose that a branch change from $(+)$ to $(-)$ has occurred at a point $\phi = \phi_0$ for a positive value of $H(\phi_0)$ (as in Fig.2) and we are interested in finding out whether another one will occur again or not. Define $\Delta^{(-)} H(\phi) = H^{(-)}(\phi) - H_{Egg}(\phi)$. If, for any $\phi$ in the Egg, $\Delta H^{(-)}$ is positive, then there will be no second hit and the two $(-)$ branch evolutions shown in Fig. 1 and in Fig. 2 can join smoothly to provide a graceful exit. If instead, $\Delta H$ vanishes somewhere in the Egg region, there will be another branch change back to the $(+)$ branch and no exit occurs. It is possible to compute $\Delta H^{(-)}$ using Eqs.(4)

$$\Delta H^{(-)}(\phi) = \int d\phi [\frac{dH^{(-)}}{d\phi} - \frac{dH}{d\phi}\Big|_{Egg}] = -\frac{1}{3}\int d\phi \Big[\epsilon(1 - \frac{1}{2}\frac{U'}{U})\Big] + \mathcal{O}(\epsilon^2)] \qquad (9)$$

Assuming, as we observed to be the case for many potentials, that $\epsilon$ remains small for a trajectory that touched the Egg (recall that $\epsilon$ is always positive), we may estimate $\Delta^{(-)} H$ by assuming that $\epsilon$ is constant

$$\Delta^{(-)} H(\phi) = -\frac{1}{3}\epsilon \int d\phi \Big[(1 - \frac{1}{2}\frac{U'}{U})\Big]$$

$$= -\frac{1}{3}\epsilon\left[(\phi - \phi_0) - \frac{1}{2}\ln\left|\frac{U(\phi)}{U(\phi_0)}\right|\right] \tag{10}$$

At the rightmost point of the Egg, $\phi_1$, $U(\phi_1) = 0$. Therefore, for any point $\phi_0$ there will be a another point, $\phi^*$, for which $\Delta^{(-)}H(\phi^*) = 0$ and also $U' > 0$. In that case the trajectory will touch the Egg again and a branch change back to the $(+)$ branch will take place. There are of course variations on this theme, possibilities of multiple hits bellow and above the Egg. We have analysed the problem numerically for a wide range of potentials and possibilities. We have not been able to construct any explicit counterexample. Although this is not an absolute no-go theorem but only a a "very-hard-to-go theorem" we are convinced that this is not the way to proceed. Rather, we suggest later what we view as a more attractive possibility for finding a solution to the graceful exit problem.

We have tried, of course, to go beyond the simplest model just illustrated in order to search for possible ways out of this conclusion. We mention here, in particular,

1. Adding classical sources in the form of a perfect fluid.

This modifies the system of equations (4) into:

$$\begin{aligned}
\dot{H} &= \pm H\sqrt{3H^2 + U + e^\phi \rho} - \frac{1}{2}U' + \frac{1}{2}e^\phi p \\
\dot{\phi} &= 3H \pm \sqrt{3H^2 + U + e^\phi \rho} \\
\dot{\rho} &+ 3H(\rho + p) = 0
\end{aligned} \tag{11}$$

These equations allow us to illustrate a point made earlier. If the potential has a minimum with vanishing cosmological constant (where $U = U' = 0$) then such point acts as an attractor for the the dilaton in the $(-)$ branch (for radiation-like matter sources, $p = 1/3\rho$). Indeed, in such case, the standard radiation-dominated cosmology

$$a = t^{1/2} \quad , \quad \phi = const. \quad , \quad \rho = a^{-4} \tag{12}$$

is a stable solution, as one can easily check by looking at small perturbations around it. By contrast, in the $(+)$ branch, nothing can keep $\phi$ from growing as long as the universe expands.

Coming back to our problem, since for a stringy fluid the ratio $p/\rho$ can be anywhere between $-1/3$ and $+1/3$, we have taken it to be some constant in this range during the Egg-hit epoch. Unfortunately, irrespectively of that ratio, equations similar to (9) and (10) still hold and force a second hit.

2. Adding extra compact dimensions.

This is a rather natural choice since we may expect that some compact dimensions evolve and eventually stabilize with string-size radii. In this case the potential is expected to depend both on the internal-dimension radii and on the effective 4-dimensional coupling, $\alpha_4 = \frac{e^\phi}{V} \equiv e^\varphi$, where $V$ is the volume of the compactified space. We assume, for simplicity, that there are three non-compact spatial dimensions with common scale factor $a$ and $n$ compact dimensions characterized by a common radius $b$. We can then define the internal analog of the Hubble parameter $H$, $K \equiv \frac{\dot{b}}{b}$. With $U = U(\varphi, b)$ and $\varphi = \phi - n \ln b$, the resulting generalization of Eqs.(4) now becomes:

$$\begin{aligned}
\dot{H} &= \pm H\sqrt{3H^2 + nK^2 + U} - \frac{1}{2}\frac{\partial U}{\partial \varphi} \\
\dot{K} &= \pm K\sqrt{3H^2 + nK^2 + U} - \frac{1}{2}\frac{\partial U}{\partial \log b} \\
\dot{\varphi} &= \dot{\phi} - nK = 3H \pm \sqrt{3H^2 + nK^2 + U}
\end{aligned} \quad (13)$$

In this case, the Egg becomes a two-dimensional surface allowing intuitively more chances for avoiding a second hit. Alas, this turns out not to be the case (at least we have found no example where graceful exit does occur). We have even tried combinations of 1) and 2), again without success.

We thus appear to be forced to the conclusion that a graceful exit from an accelerated inflation era to a standard FRW cosmology is not possible in the weak curvature regime for any sort of (semi) realistic dilaton potential. The remaining possibilities lie therefore in the strong curvature (large derivatives) regime, in which the full extent of string corrections should be felt.

It is perfectly conceivable that the strong curvature regime inducing the branch change is encountered while the string coupling is still tiny, since a constant dilaton shift is always allowed as long as the potential is negligible. Within this hypothesis, our problem becomes that of searching for conformal field theories endowed with suitable duality symmetries for describing an interesting cosmological scenario. In particular, one should select backgrounds which go to the (+) and (−) duality⊗time-reversal-related branches at large negative and positive cosmic times, respectively, with an intermediate stringy regime at small times. It is quite easy to find ad hoc modifications of Eqs.(4) with terms $O(H^4)$ and to achieve a very gracious exit this way. The challenge is to find a truly conformal background that does the same job!

We are grateful to Maurizio Gasperini, Elias Kiritsis, Costas Kounnas and Paul Steinhardt for useful and stimulating comments.